# The focusing properties of both normal and superconducting low energy CW proton Linacs*

LI Zhi-Hui (李智慧)1)

The Key Laboratory of Radiation Physics and Technology of Ministry of Education, Institute of Nuclear Science and Technology, Sichuan University, Chengdu 610065, China

**Abstract:** The continue wave (CW) high current proton linac has wide applications as the front end of the high power proton machines. The low energy part is the most difficult one and there is no widely accepted solution yet. Both normal conducting and superconducting acceleration structures are thought to be the possible solutions. Although the characteristics of normal conducting structures and superconducting ones are quite different, such as acceleration voltage, maximum electric field and so on, we found the focusing properties of the lattice composed by these two acceleration structures are quite similar for different reasons. The advantages and disadvantages of lattices composed of both the normal conducting and superconducting structures are analysed from the beam dynamics point of view, and their constraints on beam main parameters are discussed.

# 1. Introduction

Ultra high-intensity, high- energy (GeV) proton drivers are a critical technology for applications such as accelerator-driven sub-critical reactors (ADS)[1] and many HEP programs (Muon Collider) [2]. Although cyclotron [3] and FFAG [4] have great advantages in cost, it is believed that the RF linear accelerator is the most suitable type of accelerator as the high power proton driver. For the high energy part (>100MeV), the superconducting RF linac has been demonstrated in SNS [5] as one of the best choice, but for the low energy part (less than 100 or 20 MeV), there are still some debates between normal conducting and superconducting linac options. From technical point of view, the superconducting acceleration structures and normal conducting structures are quite different, but our study shows that the continue wave (CW) linac composed by the two types of acceleration structures are quite similar in beam dynamics and will be discussed in this paper.

# 2. Weak focusing in longitudinal direction

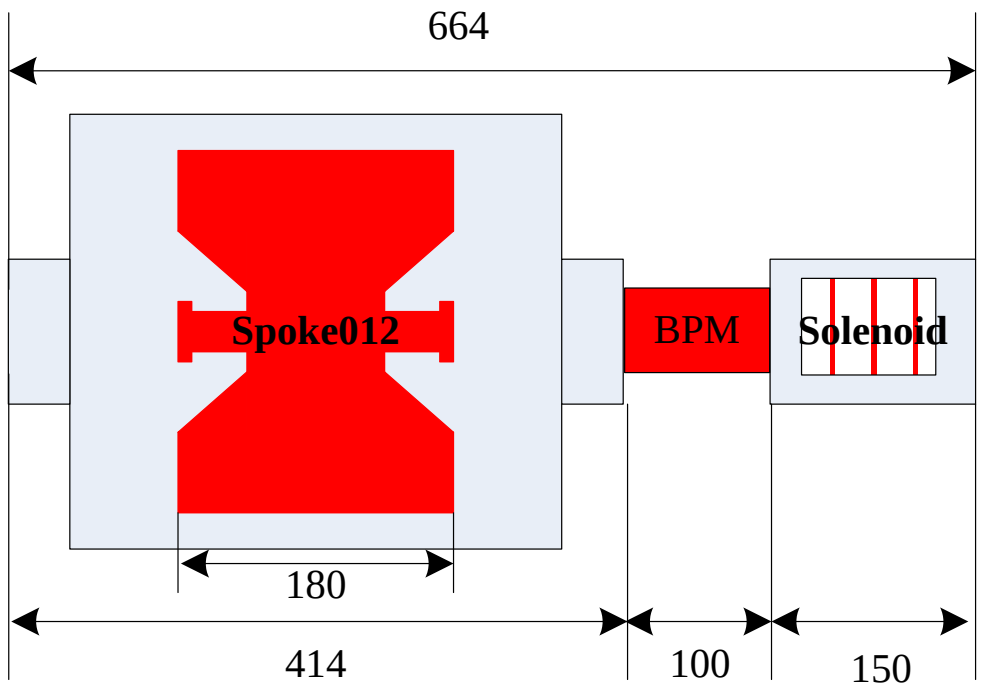

Fig.1 Lattice structure of Injector Scheme-I of CADS

One of the most significant features of low energy CW RF linear accelerator lattice is the relative weak focusing in

*Supported by the National Natural Science Foundation of China (11375122, 91126003, 11235012) and by the Strategic Priority Research Program of the Chinese Academy of Sciences，Grant No. XDA03020705）

1) E-mail:lizhihui@scu.edu.cn

longitudinal direction compared with the normal conducting pulsed one. The focusing strength of longitudinal direction can be expressed as the phase advance per meter in smooth approximation. In general the focusing strength is a function of position along beam axis, for example, for the focusing lattice showing in fig.1, the focusing lattice of Injector Scheme-I of CADS, in linear approximation, the focusing can be expressed as

$$k_{l0}(z+L_0) = k_{0l}(z) = \begin{cases} \sqrt{\frac{2\pi q E_0 T sin(-\varphi_s)}{mc^2\beta_s^3\gamma_s^3\lambda}} & |z| < l/2 \\ 0 & \beta_g\lambda/2 < |z| < L_0/2 \end{cases} \tag{1}$$

where $l = 110\,mm$ is the effective length of the Spoke012 cavity and $L_0 = 664\,mm$ is the period length of the focusing lattice, and the longitudinal motion can be expressed as

$$\varphi'' + k_{l0}^2(z)\varphi = 0 \tag{2}$$

In smooth approximation, the zero current phase advance per period can be expressed as equation (3) [6]

$$\cos\sigma_0 = cos\theta - \frac{1}{2}\frac{L_0-l}{l}sin\theta \tag{3}$$

where $\theta = k_{0l}l$ is the focusing strength parameter. The relation between $\sigma_0$ and θ is shown in fig.2.

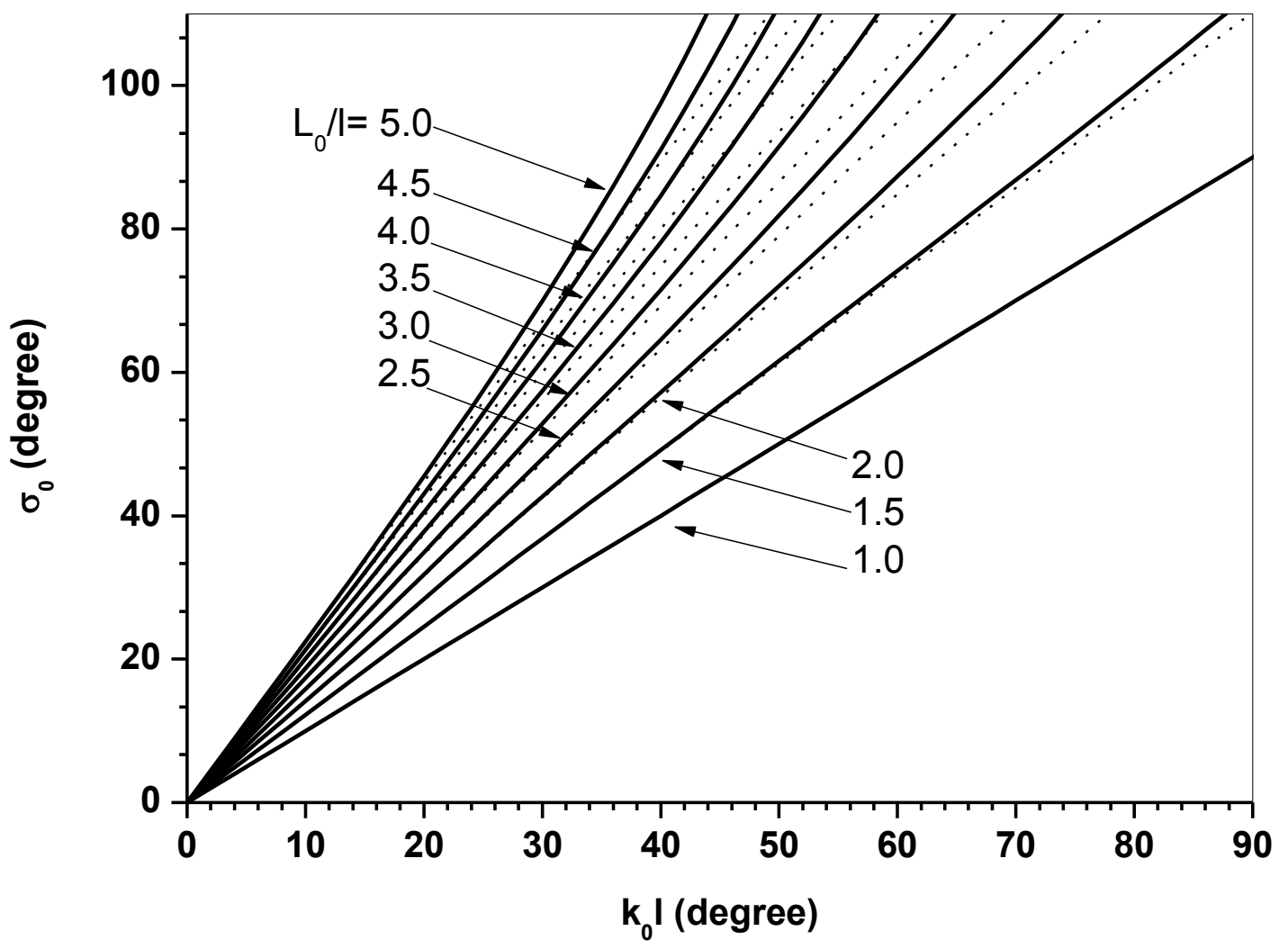

Fig.2 zero current phase advance as function of the focusing strength parameter of RF cavity, the solid lines are calculated from equation (3) and the dot lines are calculated from equation (4)

The relation between phase advance and the $k_{l0}$ is rather complicated in equation (3), usually for small $\sigma_0$, by expanding the triangle functions in equation(3) as power serious of $\sigma_0$ and θ, then we can get a simply relation as equation (4) shows,

$$\sigma_0^2 \approx k_{l0}^2 L_0 l = \frac{2\pi q V_0 sin(-\varphi_s)}{mc^2\beta_s^2\gamma_s^3}\frac{L_0}{\beta_s\lambda} \tag{4}$$

where $V_0$ is the effective voltage and it is also plotted in fig.2 as dot lines. We can see for small phase advance, it is accurate for small $L_0/l$, but as $L_0/l$ increase, equation (4) get a small phase advance as that get from equation (3). For the CW normal conducting structures, the average power deposited on the cavity surface is at least several hundred times that on the pulsed cavity surface, how to transfer the heat from the cavity surface to make sure that cavity will not be ruined with too high temperature is one of the critical issues in cavity design. Since the power is proportional to the square of the effective voltage, the field level in the CW acceleration structures is only 1/2 or 1/3 as that in the pulsed ones in order to keep the power density in a reasonable level, so the average acceleration gradient of the low energy CW normal conducting Linac is only about 1MV/m. The longitudinal focusing strength is thus quite weak compared with the pulsed one.

For the superconducting cavities, because of the nearly zero surface resistance, the field level in the cavity is usually much higher than that in normal conducting cavities and the acceleration gradient of the cavity can reach to 4 to 8 MV/m for low energy superconducting cavities. However, the focusing strength is not only determined by the effective voltage, but also determined by the period length, that means the focusing is determined by the average acceleration gradient of the linac. For the superconducting cavities, the existence of the static magnetic field will increase the surface resistance and may cause it to quench, the cavity needs to be well screened from any static magnetic field, which makes it impossible to integrate the transverse focusing lens with the cavity just as the normal conducting Alvarez DTL cavity does. As a consequence, the focusing period length will be much larger than the normal conducting one, especially at the low energy part, where the space charge effect is important and transverse focusing has to be applied between every cavity. Besides this, the zero current phase advance per period should not exceed 90 degree in order to make the beam dynamical stable. That means in some cases, even the cavity can provided higher voltage, the acceleration potential of the cavities cannot be fully used. All this makes the longitudinal focusing strength of the low energy CW proton linac is weaker compared with the pulsed normal conducting one and is comparable with the CW normal conducting one.

# 3. Balanced focusing in transverse and longitudinal directions

As discussed in the proceeding section, the weak longitudinal focusing is one of the natural features of the CW low energy linac, no matter it is composed by normal conducting cavities or superconducting ones. How about the transverse focusing? Is there any correlation between longitudinal and transverse focusing? Our study shows that for high current machine, there are some correlations between transverse and longitudinal focusing. Fig.2 shows the emittance growth for a 20 mA, 10 MeV proton beam transporting throw a focusing channel with different focusing strengths in longitudinal and transverse directions. The particle initial distribution is 3σ truncated Gaussian distribution in six dimensional phase space and initial RMS transverse and longitudinal emittances are 0.24 and 0.27 mm.mrad, respectively. The focusing is composed by 74 periods and each period is composed by a RF gap and a doublet, and the period length is 760 mm, just the same as the China ADS Injector Scheme I former design. The beam transport properties are simulated by Tracewin [7] code with 2 dimensional PIC space charge routine.

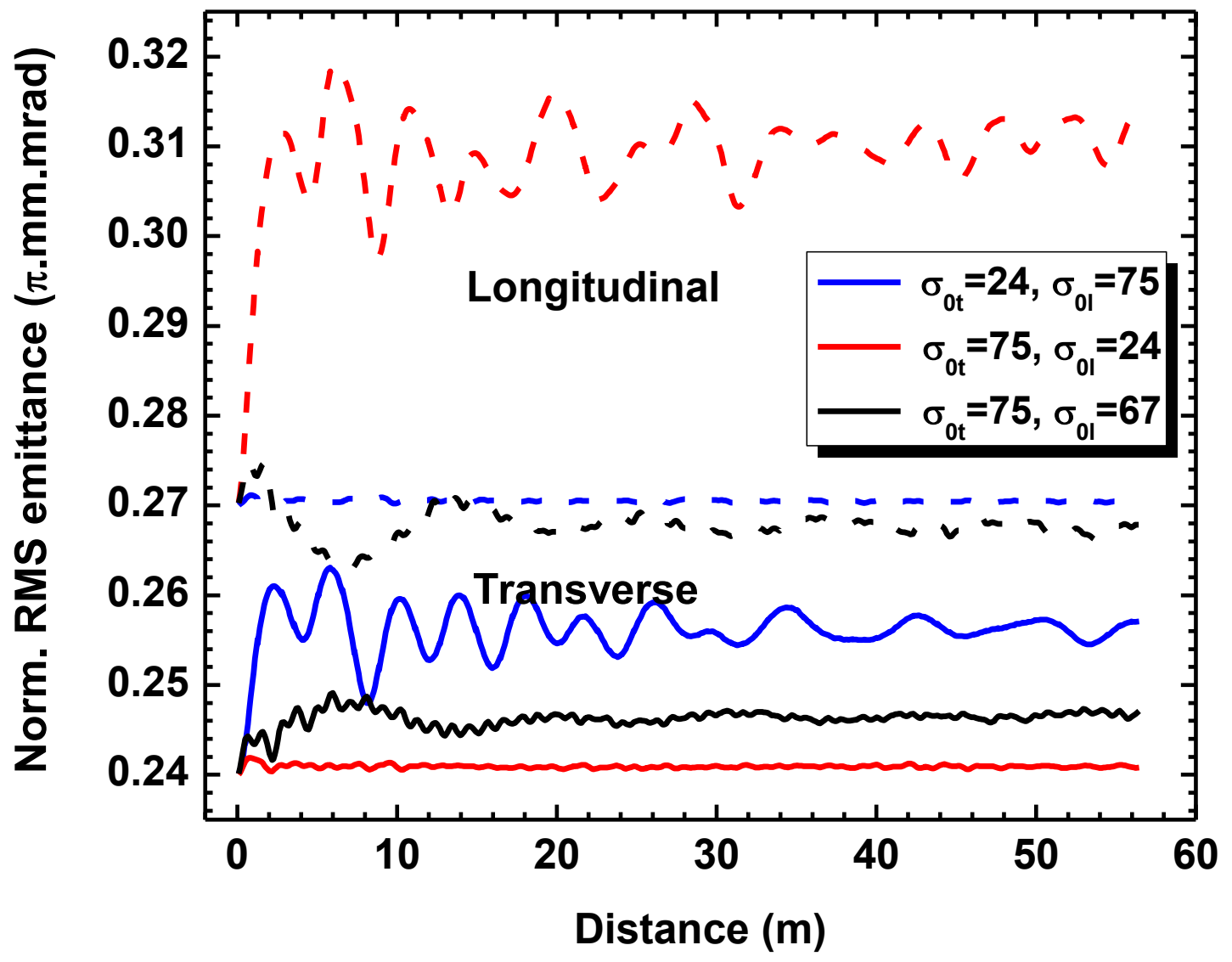

Fig. 3. Emittance growth (dash line: longitudinal emittances; solid lines: transverse emittances)

The blank lines in Fig. 3 shows the emittance growth with transverse and longitudinal phase advance per period 75 degree and 67 degree, which satisfy the equipartitioning condition and the emittance growths in longitudinal and transverse plans are almost zero. The red and blue lines show the emittance growths for the case that one of the two directions is weakly focused compared with the other one, but we still make sure that the working points are located in the resonance free region in Hofmann Chart [8]. We can see significant emittance growth in the weak focusing plane for both cases, it seems that for the high current machine it is better to give balanced focusing in each directions.

In order to study the reason for the emittance growth, we studied the emittance growth with different particle distributions and the results are showed in Fig. 4. The beam current and energy is just the same as the case in Fig.3 and the phase advance in transverse and longitudinal directions are 24 and 75 degree, respectively. We can see the transverse emittance growth is closely dependent on the distribution type, for the uniform beam, the emittance growth is almost zero as blue and pink lines showing. It indicates that it is an effect of nonlinear space charge. We can see at first several periods the longitudinal emittance oscillates strongly and then gradually approaches to a stable value and looks very similar with the charge redistribution. Another clue comes from the fact that the tune depression of the weak plane is very small, only about 0.4, and some space charge connected resonance may be exited and causes the emittance growth. So in order to decrease the emittance growth, it is better to set the focusing strength in transverse and longitudinal as close as possible. Together with the equipartitioning condition [9], we can deduce that the emittance ratios between different plans should be close to 1 as possible.

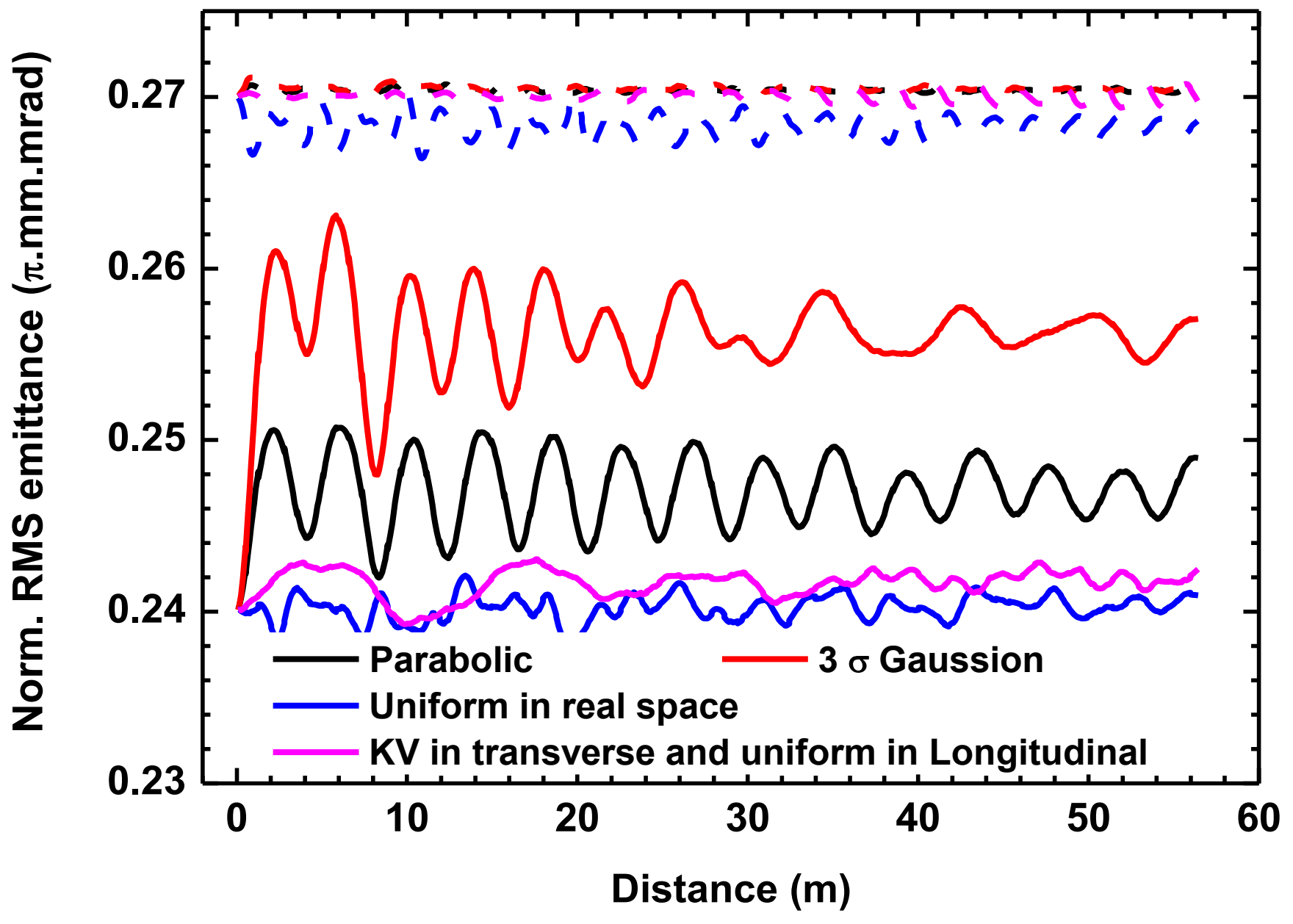

Fig. 3. Emittance growth for different initial distributions. The phase advances per period for transverse and longitudinal directions are 24 and 75 degree, respectively.

# 4. Current limits

As front end of a high power machine, the capacity of the beam current is one of the most important parameters. As discussed, the transverse and longitudinal focusing are weak for both normal and superconducting machine, so the maximum beam current should be limited as expected. Usually the maximum beam current is determined by the tune depression. For high current machine, the tune depression should not be too small. In order to keep the machine in stable operation, the tune depression should satisfy the following relation [10],

$$1 \geq \mu \geq 0.5 \tag{5}$$

Fig. 5 shows the tune depression as a function of beam current with different energy when transporting throw the focusing channel. The zero current phase advances per period in transverse and longitudinal directions are set as 75 and 67 degree respectively and the lattice structure is just the same as that in previous section. We can see, for 3 MeV proton beams, when beam current is increased to 35 mA, the longitudinal tune depression is decreased to about 0.5. For the real machine, the matching between different sections cannot be perfect, the beam parameters cannot be precisely measured, and the errors for all parameters setting to all kinds of elements inevitably exist, so beam current capacity will be much smaller than this number. Fig.5 also indicates that the low energy part is the bottle neck of high beam current.

For the normal conducting structures, as the beam current increase, the phase advance will decrease and the beam radius will increase as equation (6) shows [10],

$$\bar{R} = \sqrt{\frac{\varepsilon L_0}{\sigma}} \tag{6}$$

Suppose the minimum tune depression is 0.5, then the beam radius will be about 1.4 time that of zero current. It implies that for high current beam machine, the aperture should be bigger, which will further decrease the shunt impedance and increase the power density at the given acceleration gradient, and this may require to decrease the acceleration gradient and decrease the longitudinal focusing strength further.

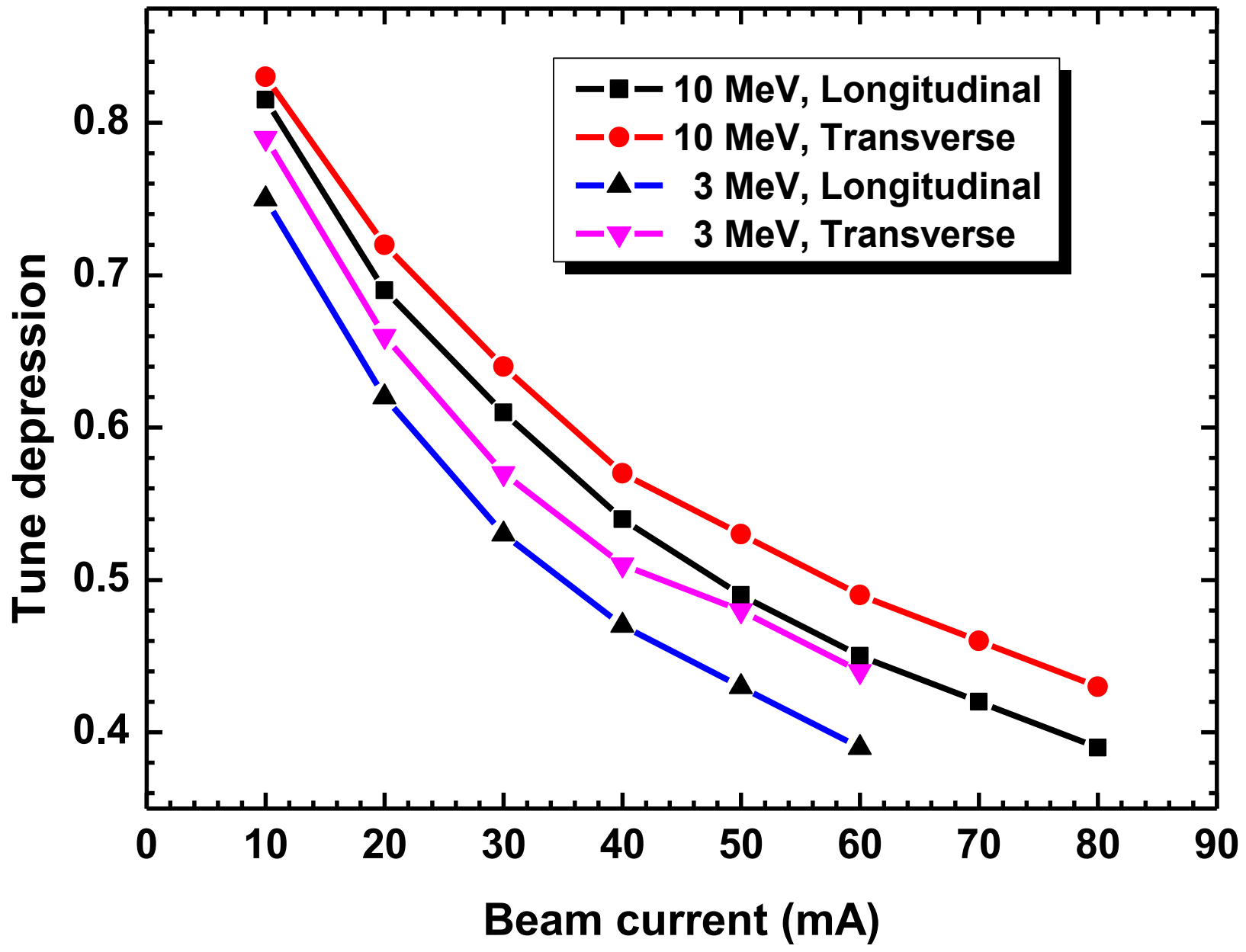

Fig. 5: Tune depression as function of beam current

# 5. Requirements for the acceleration structures

Since both transverse and longitudinal focusing is weak and the beam is space charge dominated when beam current is greater 30 mA, it is natural to ask how to increase the focusing strength. For the normal conducting structure, it is obvious that the βλ/2 structure is better since it can produce more high acceleration gradient, such as CH structure. Furthermore, because the period length is longer, about 7-10 βλ, it is possible to keep the transverse focusing elements outside the cavity, so that the shunt impedance can be increased. The other important thing is to determining the condition in which the cavity can stably work in CW mode, for example, the maximum surface electric field and maximum power density, all these should be determined experimentally.

For the superconducting structures, the optimization should be focused on how to increase the filling factor (the ratio between effective acceleration length and the period length) of the cavity. For the low energy structure, the acceleration gradient is no longer important parameters and 4-6 MV/m is already enough, the more important thing is to develop some kind of compact structure, so that the acceleration potential can be fully explored.

# 6. Conclusions

Because of the low field level in the normal conducting CW cavities and long period length of the superconducting linac lattice, the longitudinal and transverse focusing is weak for the low energy CW proton linac. The requirements of balanced focusing between different directions and equipartitioning condition ask that the emittance of all the directions should be as closer as possible. The weak focusing features determines that the maximum beam current of CW low energy linac is limited and for safe and stable operation, it may be better to keep the beam current less than 30 mA, unless some kinds of new structures are invented.